\documentclass{article}
\usepackage{amsmath,amssymb}
\usepackage{mathtools}
\usepackage{cite}
\usepackage{verbatim}
\usepackage{float}
\usepackage{subcaption}
\usepackage{amsfonts}
\usepackage[utf8]{inputenc}
\usepackage{graphicx}
\newcommand{\be}{\begin{equation}}
\newcommand{\ee}{\end{equation}}
\newcommand{\bea}{\begin{eqnarray}}
\newcommand{\eea}{\end{eqnarray}}
\newcommand{\bean}{\begin{eqnarray*}}
\newcommand{\eean}{\end{eqnarray*}}

\def\beq{\begin{equation}}

\def\eeq{\end{equation}}
\def\R{\mathcal{R}}

\relax

\def\d{\partial}

\begin{document}
\begin{titlepage}
\begin{center}

\vskip 20mm

{\Huge WKB method and quasinormal modes of string-theoretical d-dimensional black
holes}

\vskip 10mm

Filipe Moura$^{\dag}$ and Jo\~ao Rodrigues$^{\ddag}$

\vskip 4mm

\emph{$^{\dag}$Departamento de Matem\'atica, Escola de Tecnologias e Arquitetura and \\Instituto de Telecomunica\c c\~oes, \\ISCTE - Instituto Universit\'ario de Lisboa,
\\Av. das For\c cas Armadas, 1649-026 Lisboa, Portugal}
\vskip 2 mm
\emph{$^{\ddag}$Centro de An\'alise Matem\'atica, Geometria e Sistemas Din\^amicos,\\ Departamento de Matem\'atica,\\ Instituto Superior T\'ecnico,\\
Av. Rovisco Pais, 1049-001 Lisboa, Portugal}

\vskip 4mm
\texttt{fmoura@lx.it.pt}, \quad
\texttt{joao.carlos.rodrigues@tecnico.ulisboa.pt}

\vskip 6mm

\end{center}

\vskip .2in

\begin{center} {\bf Abstract } \end{center}
\begin{quotation}\noindent
After a brief introduction to quasinormal modes in dissipative systems, we review the WKB formalism in the context of the analytical calculation of quasinormal frequencies. We apply these results to the calculation of quasinormal frequencies associated with gravitational perturbations of d-dimensional spherically symmetric black holes with string corrections. We do this for two distinct limits: the eikonal limit and the asymptotic limit.

\end{quotation}

\vfill


\end{titlepage}

\eject

\newpage
\section{Introduction and summary}
\noindent

Gravitational wave detectors can directly measure quasinormal ringing frequencies, which carry unique information about parameters of the black hole in the ringdown phase resulting from a binary black hole coalescence. This feature turns quasinormal modes into preferential probes for testing theories of gravity beyond Einstein, since the ringing frequencies represent a universal part of the gravitational wave signals. With the advent of gravitational wave astronomy, therefore, interest in the study of black hole quasinormal modes has raised, as their measurement can provide a test to modified gravity theories \cite{branes}.

We provide an introductory review to this topic of great current interest. In section \ref{2} we start by introducing quasinormal modes as solutions to field equations of dissipative systems, and we verify that quasinormal frequencies arise as the poles of a frequency domain Green’s function. We then define black hole quasinormal modes and the boundary conditions they must verify. In section \ref{wkb} we briefly review WKB theory in order to obtain approximations to general solutions of Schr\"odinger--like differential equations, like the ones describing black hole perturbations. In section \ref{cmp} we introduce a $d$-dimensional spherically symmetric black hole solution with leading string corrections obtained by Callan, Myers and Perry, and briefly review the formalism for gravitational perturbations applied to it. We then compute analytical expressions for quasinormal frequencies, associated with tensorial gravitational perturbations of the black hole solution we introduced. We do this for two distinct limits. In section \ref{qnmeik}, we compute these frequencies in the eikonal limit, resorting to the WKB approximations we addressed. Finally in section \ref{qnmd} we compute these frequencies in the asymptotic highly damped limit, studying two different monodromies of the perturbation, when analytically continued to the complex $r$-plane.

\section{Quasinormal modes}
\label{2}
\subsection{Dissipative systems and quasinormal modes}
\label{dis}
\noindent

We start with a digression on physical systems described by differential equations whose spatial variables are defined on non compact subsets of $\mathbb{R}^n$. Here we closely follow \cite{stars}, conveniently adapted to our context. As an example, let us picture a general 1-dimensional physical system, defined by the wave equation
\begin{equation}
    \frac{\partial^2 \psi}{\partial t^2} + \left(V(x) - \frac{\partial^2 }{\partial x^2}\right) \psi = 0
    \label{12}
\end{equation}
where $V$ is a continuous positive function of compact support in $\mathbb{R}$, meaning that
\begin{equation}
    V(x) = 0
\end{equation}
for all $x \in \mathbb{R}$ such that $|x| > L$ with $L \in \mathbb{R}^+$. Let us choose general initial data of compact support
\bea
    \psi(x,0) = \Gamma(x), \\
    \frac{\partial \psi }{\partial t}(x,0) = \Omega(x)
\eea
for some $\Gamma \in C^2(\mathbb{R})$ and $\Omega \in C^1(\mathbb{R})$.

Now, we consider the unique solution $\psi$ of the differential equation $(\ref{12})$, with initial data as above. The Laplace transform of this solution takes the form
 \begin{equation}
     \mathcal{L}\left(\psi\right)(x,s) = \int_0^{+\infty}e^{-st}\psi(x,t)dt.
 \end{equation}
The differential equation satisfied by $\mathcal{L}(\psi)$ is
\begin{equation}
    s^2\mathcal{L}(\psi) - \frac{\partial^2\mathcal{L}}{\partial x^2}(\psi) + V \mathcal{L}(\psi) = s\Gamma + \Omega.
    \label{13}
\end{equation}
Because we are working with initial data of compact support, we know $\psi$ is bounded \cite{stars}. Hence, the Laplace transform $\mathcal{L}(\psi)$ is analytic for $s \in\hspace{2pt} ]0,+\infty[$ and admits an analytic continuation \cite{stars} onto the complex half plane $\Re (s) > 0$ \footnote{We denote by $\Re (s)$/$\Im (s)$ the real/imaginary parts of the complex number $s.$}.

Now, let us write the Green function of the differential equation $(\ref{13})$. Such function, takes the well known form
\begin{equation}
    G(s,x,x') = \begin{dcases}
    \frac{f_-(s,x')f_+(s,x)}{W(s)} \hspace{3pt}; \hspace{3pt} x' \le x \\
     \frac{f_-(s,x)f_+(s,x')}{W(s)} \hspace{3pt}; \hspace{3pt} x' > x
    \end{dcases}
\end{equation}
where $f_-$ and $f_+$ are two linearly independent solutions of the homogeneous differential equation, associated with $(\ref{13})$, and $W$ denotes the respective Wronskian. In order to simplify the notation, we define the function
\begin{equation}
    \Lambda \coloneqq s\Gamma + \Omega.
\end{equation}
Using the previous definition, we can write the particular solution of the differential equation $(\ref{13})$ as
\begin{equation}
    \mathcal{L}(\psi)(s,x) = \int_{-\infty}^{+\infty}G(s,x,x')\Lambda(s,x')dx'.
    \label{14}
\end{equation}

Considering the Laplace transform definition, we know $\mathcal{L}(\psi)$ is bounded as a function of $x$. Indeed, since $\psi$ is bounded, we know that $|\psi|$ has a majorant $M \in \mathbb{R}^+$. Thus, for some $s$ in the complex half plane $\Re(s) > 0$, we have
\begin{equation}
    \left\vert\mathcal{L}(\psi)\right\vert = \left\vert\int_0^{+\infty}e^{-st} \psi(x,t) dt\right\vert \le \int_0^{+\infty}\left\vert e^{-st}\right\vert\left\vert\psi(x,t)\right\vert dt \le M \int_0^{+\infty}e^{- \Re(s)t }dt = \frac{M}{\Re(s)}.
\end{equation}
Moreover, for a fixed $s$ in the complex half plane $\Re(s) >0$ and for $|x| > L$, the general solution of the homogeneous differential equation, associated with $(\ref{13})$, is a linear combination of the functions $e^{\pm xs}$. Since $\mathcal{L}(\psi)$ is bounded as a function of $x$, we are forced to choose a Green function $G$ such that
\bea
    f_+(s,x) \propto e^{-sx} \hspace{3pt};\hspace{3pt} x \ge L, \\
    f_-(s,x) \propto e^{sx} \hspace{3pt};\hspace{3pt} x \le -L.
\eea
The above conditions uniquely define the Green function of $(\ref{13})$ and the respective particular solution through $(\ref{14})$.

It is within this context that quasinormal frequencies emerge. Quasinormal frequencies are defined as the complex numbers $s_n \in \mathbb{C}$ such that
\begin{equation}
    f_+(s_n,x) = c(s_n) f_-(s_n,x)
    \label{156}
\end{equation}
for all $x \in \mathbb{R}$, where $c(s_n) \in \mathbb{C}$. For these values of $s$, the solutions $f_+$ and $f_-$ become linearly dependent, with respect to $x$. When this happens, the Wronskian $W$ vanishes and the chosen Green function becomes singular. The corresponding functions $f_-(s_n,x)$ and $f_+(s_n,x)$ are called quasinormal modes. We did not prove that such frequencies exist, but in fact they do. In \cite{existencia}, for the conditions we are considering, a countable family of such frequencies, in the complex half plane $\Re(s) < 0$, was proven to exist. We know $\mathcal{L}(\psi)$ admits an analytic continuation onto the complex half plane $\Re(s) > 0$. Thus, there should not exist any quasinormal frequency with positive real component.

\begin{figure}[h]
\centering
\includegraphics[width=0.5\textwidth]{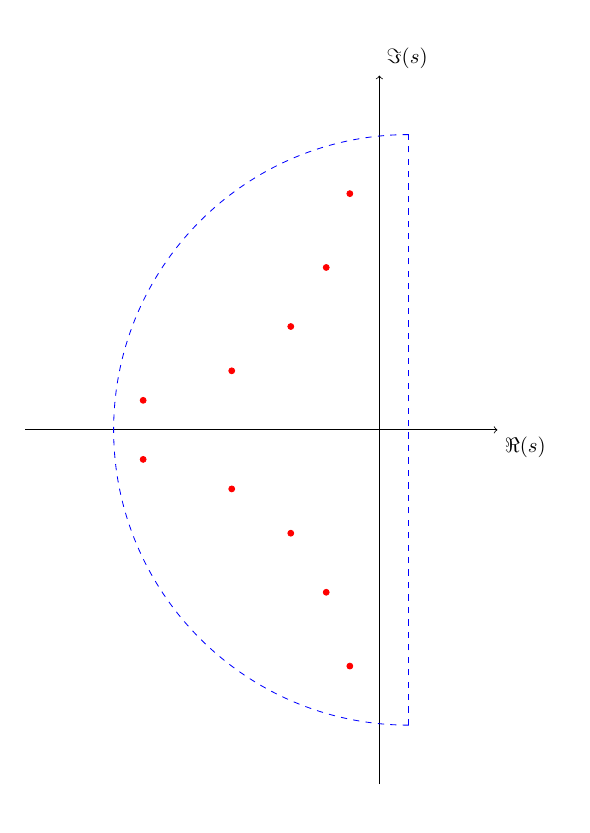}
\caption{Pictorial representation of the closed contour needed to compute the integral over $s$ in (\ref{15}) as the blue dashed line. Furthermore, possible quasinormal frequencies are depicted by conjugated red dots. These conjugation pairs arise from conjugation of (\ref{156}).}
\label{fig2}
\end{figure}

The quasinormal frequencies and quasinormal modes come into play when we reconsider the solution to the original differential equation $(\ref{12})$. Indeed, we can use Mellin's inverse formula to write $\psi$ as
\begin{equation}
    \psi(x,t) = \frac{1}{2\pi i }\int_ {a -i\infty}^{a+i\infty}e^{st}\mathcal{L}(\psi)(s,x)ds = \frac{1}{2\pi i}\int_{a-i\infty}^{a+i\infty}e^{st}\int_{-\infty}^{+\infty} G(s,x,x')\Lambda(s,x')dx'ds
    \label{15}
\end{equation}
for some $a >0$. Now, we can compute the integral over $s$, using the contour depicted in figure \ref{fig2}. To do this, we define
\begin{equation}
    \mathcal{F}(s,x) \coloneqq \frac{e^{st}}{2\pi i}\int_{-\infty}^{+\infty} G(s,x,x')\Lambda(s,x')dx'
\end{equation}
and rewrite (\ref{15}) as
\begin{equation}
    \psi(x,t) = \int_{a-i\infty}^{a+i\infty}\mathcal{F}(s,x)ds.
    \label{16}
\end{equation}
Because we chose initial data of compact support, we know $\Lambda$ is compactly supported on $x'$ as well. Therefore,
there can be no singularities of $\mathcal{F}$, arising from the integration. On the other hand, we know $G$ is singular for quasinormal frequencies $s_n$. As our contour encloses these frequencies, the residue theorem yields
\begin{equation}
    \oint_\mathcal{D}\mathcal{F}(s,x) ds = \int_{a-i\infty}^{a+i\infty}\mathcal{F}(s,x)ds + \int_{C}\mathcal{F}(s,x)ds = 2\pi i\sum_{n \in I}\text{Res}\left(\mathcal{F},s_n\right)(x)
\end{equation}
where we denoted the blue contour and the associated arc shaped portion by $\mathcal{D}$ and $\mathcal{C}$ respectively. Furthermore, we denoted the countable index set of quasinormal frequencies by $I$. Rewriting the equation above, using $(\ref{16})$, yields
\begin{equation}
    \psi(x,t) = 2\pi i\sum_{n \in I}\text{Res}\left(\mathcal{F},s_n\right)(x) - \int_{C}\mathcal{F}(s,x)ds.
    \label{187}
\end{equation}
We arrived at an expression relating $\psi$ to the quasinormal frequencies.

The second contribution to $\psi$, in the equation above, is due to the contour integration along $\mathcal{C}$. We notice that $\Re(s) \ll -1$ in most of $\mathcal{C}$. Thus, the exponential factor in the definition of $\mathcal{F}$ is bound to decay rapidly as $t$ increases and so is the contribution as a whole \cite{latetime}. This renders the contribution negligible for the late time behaviour of the system. Hence, we end up with the asymptotic behaviour
\begin{equation}
    \psi(x,t) \sim 2\pi i \sum_{n\in I}\text{Res}\left(\mathcal{F},s_n\right)
\end{equation}
in this limit. Finally, the expression above can be recast in the form
\begin{equation}
    \psi\left(x,t\right) \sim \sum_{n\in I} C_n e^{s_nt}f_+\left(s_n,x\right)
    \label{22}
\end{equation}
for finite values of $x$ and for some $C_n \in \mathbb{C}$ \cite{stars}.

As we can see from $(\ref{22})$, the unique solution of our physical problem will decay exponentially fast in time, for spatially bounded regions. In this sense, equations such as $(\ref{12})$, defined on non compact spatial domains, model physical dissipative systems. Solutions to these systems, as we just saw, cannot be fully described by normal modes. Instead, in the late time behaviour of the system, they can be well approximated by quasinormal modes. In this way, they play the analogous role of normal modes in systems of compact spatial domain. A major distinction between the two sides is that normal frequencies are real while quasinormal frequencies may be complex.

\subsection{Quasinormal modes of black holes}
\noindent

A general static spherically symmetric metric in $d$ dimensions has the form
\begin{equation}
    ds^2 = -f dt \otimes dt + \frac{1}{f}dr\otimes dr + r^2d^2\Omega_{d-2},
    \label{35}
 \end{equation}
where $d\,\Omega^2_{d-2}$ is the canonical metric tensor field of the unit ($d$-2)-sphere.

We consider a complex massless scalar test field $\psi$, minimally coupled to a background of the form (\ref{35}).
In such background, this scalar field can be decomposed as \cite{branes}
\begin{equation}
   \psi(t,r,\theta,\phi) = \sum_{l=0}^{+\infty}\sum_ {m=-l}^l  \frac{e^{i\omega t}}{r}\psi_r(r)  Y_{lm}(\theta,\phi).
   \label{4}
\end{equation}
Each component $\psi_r$ obeys a second order field equation of the form
\begin{equation}
    \frac{d^2\psi_r}{dx^2} + \left(\omega^2 - V\right) \psi_r = 0
    \label{36}
\end{equation}
where $x$ is the tortoise coordinate defined as
\begin{equation}
    dx := \frac{dr}{f}.
    \label{50}
\end{equation}
The potential $V$ in (\ref{36}) can be expressed, with respect to $r$, as
\begin{equation}
     V[f(r)] = f(r)\left(\frac{l(l+d-3)}{r^2} + \frac{(d-2)(d-4)f(r)}{4r^2} + \frac{(d-2)f'(r)}{2r}\right).
     \label{153}
\end{equation}
This potential (\ref{153}) is valid for minimally coupled massless scalar fields in the background of any metric of the type (\ref{35}), in Einstein gravity, but also in the presence of higher derivative corrections (which appear implicitly through the $\lambda'$ corrections of $f(r)$).

Now we admit the background metric (\ref{35}) to be a black hole with horizon radius $R_h$, and we concern ourselves with boundary conditions associated to (\ref{36}). We are interested in a solution of the perturbation $\psi$, propagating in the spatial region outside of the black hole. That is to say, we only seek a solution for values of $r$ in the region $r>R_h$. The tortoise coordinate $x$, defined as we did, takes values in $\mathbb{R}$ and has the asymptotic behaviours $x \to + \infty$ when $r \to +\infty$ and $x \to -\infty$ when $r \to R_h^+.$ It is also easy to see that
\begin{equation}
    \lim_{r \to +\infty} V[f(r)] = 0,\, \lim_{r \to R_h^+}V[f(r)] = 0.
    \label{19}
\end{equation}
From the limits above, we gather that, in the boundaries of the region we are interested in, the solutions to the master equation (\ref{36}) behave as
\begin{equation}
    \psi_r(x) = A_+e^{i\omega x} + A_- e^{-i\omega x} \label{ho}
\end{equation}
for some $A_\pm\in \mathbb{C}$. Because nothing should leave the event horizon, we impose the boundary condition
\begin{equation}
    \psi_r(x) \propto  e^{i\omega x}, \, x \to -\infty.
    \label{20}
\end{equation}
On the other hand, as we are describing an isolated system, we want to disregard unphysical waves coming from spacial infinity. Thus, we impose the boundary condition
\begin{equation}
    \psi_r(x) \propto e^{-i\omega x}, \, x \to +\infty.
    \label{21}
\end{equation}

Conditions $(\ref{20})$ and $(\ref{21})$, independent of initial data, define the asymptotic behaviour of the dominant radial contribution to $\psi$ in the late time behaviour of the system: the \emph{quasinormal modes}. These modes are defined as the solutions of equation (\ref{36}), obeying the boundary conditions above for some associated quasinormal frequency $\omega$.

In section \ref{dis}, we defined quasinormal frequencies as values of $s$. However, it is common to write the variable associated with the Laplace transform as $s = i\omega.$ Thus, quasinormal frequencies are usually defined as values of $\omega$ instead. Using this notation, it is easy to see that our latest definitions of quasinormal modes and frequencies match those previously introduced.

As an observer, far away from the black hole, we are interested in the limit of spatial infinity, for the late time behaviour of $\psi$. Looking at $(\ref{22})$, we see that in order to know $\psi$, we first need to know the quasinormal frequencies and the associated complex constants $C_n$, also called quasinormal excitation coefficients. How do we compute these? Regarding the excitation coefficients, the only way to compute them is by providing the problem with initial data \cite{branes}. On the other hand, quasinormal frequencies only depend on the metric tensor field of the background black hole space time.

We are left with the task of computing quasinormal frequencies. In general, it is very complicated (if not impossible) to obtain an analytical expressions of them. However, there are two limiting cases where we might have an easier time looking for analytical expressions. These limiting cases target existing quasinormal frequencies with, operationally wise, desirable properties. The first limiting case targets quasinormal frequencies obeying the master equation for arbitrarily large values of the azimuthal number $l$ \footnote{In our example, this number arises from the spherical harmonic decomposition $(\ref{4})$. Naturally, this limiting case is restricted to background black hole space times whose metric tensor field allows for similar decompositions.}. This limit is usually called the eikonal limit. The second limiting case targets quasinormal frequencies $\omega$ whose imaginary part is much larger (in magnitude) than the real part: $|\Im(\omega)|\gg |\Re(\omega)|$. This limit is usually called the asymptotic limit. We will be working with both these limits.

Finally, while we postpone to later sections the introduction and discussion of actual analytical methods to compute quasinormal frequencies in both limiting cases introduced, we take one last note. In section \ref{dis}, we saw that quasinormal frequencies were restricted to the complex half plane $\Re(s) < 0$. It is clear that this restriction translates to $\Im(\omega)>0$ in the new notation.

\section{WKB theory: a review}
\label{wkb}
\noindent

We briefly review the WKB theory and explain some basic results which will be used extensively. The name WKB comes from the founders Gregor Wentzel, Hendrik Anthony Kramers, and Léon Brillouin.

\subsection{Defining purpose and Schr\"odinger like equations}
\noindent

The broad purpose of the WKB theory is to provide approximations to general solutions of differential equations whose highest order derivative is
multiplied by some small real parameter. In this work, we will only deal with ordinary homogeneous linear differential equations. Hence, we restrict
ourselves to this class of differential equations. Let us write a general differential equation of this class as
\begin{equation}
    \epsilon \frac{d^ny}{dx^n} + \sum_{m=0}^{n-1} a_i\frac{d^my}{dx^m}= 0
\end{equation}
where $a_i \in C^k(\mathbb{R})$, for some $k \in \mathbb{N}$. The WKB method proposes an asymptotic series expansion of the form
\begin{equation}
    y\sim \prod_{n=0}^{+\infty}e^{S_n\delta^{n-1}}.
    \label{23}
\end{equation}
where $S_n \in C^{k_n}(\mathbb{R})$ for some $k_n \in \mathbb{N}$ and where $\delta$ is a small real parameter. The scaling of $\delta$ with $\epsilon$ is determined by substitution along with the functions $S_n$.

As an example, we consider the Schr\"odinger like equation
\begin{equation}
    \epsilon^2\frac{d^2 y}{dx^2} = Qy
    \label{24}
\end{equation}
where $\epsilon^2$ is a small real parameter and $Q \in C^{\infty}(\mathbb{R})$. Replacing the asymptotic series $(\ref{23})$ in the differential
equation above yields
\begin{equation}
    \epsilon^2\left[\frac{1}{\delta^2}\left(\sum_{n=0}^{+\infty}\delta^n\frac{dS_n}{dx}\right)^2 + \frac{1}{\delta}\sum_{n=0}
     ^{+\infty}\delta^n \frac{d^2 S_n}{dx^2}\right] = Q.
\end{equation}
We fix $\delta = \epsilon$ and proceed to solve the differential equation perturbatively in powers of $\epsilon$. The equation of leading order is
\begin{equation}
   \left(\frac{d S_0}{dx}\right)^2 = Q
\end{equation}
whose solutions are (up to an additive constant, which we will ignore since $(\ref{24})$ is linear)
\begin{equation}
    S_0(x) = \pm \int \sqrt{Q(x)}dx.
\end{equation}

The differential equation of first order in $\epsilon$ is
\begin{equation}
    2 \frac{d S_0}{dx}\frac{dS_1}{dx} + \frac{d^2S_0}{dx^2}=0
\end{equation}
whose solution is (also up to an additive constant)

\begin{equation}
    S_1(x) = -\frac{1}{4}\log\left(Q(x)\right).
\end{equation}
We can follow this procedure iteratively, solving differential equations of higher order in $\epsilon$. These equations take the generic form (for
$n \geq 2$)
\begin{equation}
    2\frac{dS_0}{dx}\frac{dS_n}{dx} + \frac{d^2 S_{n-1}}{dx^2} + \sum_{j=1}^{n-1}\frac{dS_j}{dx}\frac{dS_{n-j}}{dx}=0.
\end{equation}
Because $(\ref{23})$ is an asymptotic series, there is usually an accuracy limit associated as a representation of the solution to the differential equation considered. Hence, only a finite amount of functions $S_n$ are of interest to us.

Up to zeroth order in $\epsilon$, the general solution of the differential equation $(\ref{24})$ can be expressed as
\begin{equation}
    y(x) \sim \frac{\mathcal{C}_+}{\sqrt[4]{Q(x)}}\exp\left(\frac{1}{\epsilon}\int\sqrt{Q(x)}dx\right) + \frac{\mathcal{C}_-}{\sqrt[4]{Q(x)}}\exp\left(-\frac{1}{\epsilon}\int\sqrt{Q(x)}dx\right)
    \label{25}
\end{equation}
for some $\mathcal{C}_\pm \in \mathbb{C}$.

\subsection{Analytic continuation and Stokes lines}
\noindent

Now, we study the validity domain of the solution $(\ref{25})$, when analytically continued to complex $x$-plane. In order to do that, we rederive the solution $(\ref{25})$, using a different approach. First, let us assume $Q$ can be analytically continued onto the complex $x$-plane. Furthermore, let us rewrite equation $(\ref{24})$ as
\begin{equation}
    \frac{d^2y}{dx^2} - \frac{Q}{\epsilon^2}y = 0.
\end{equation}
We introduce the new dependent variable
\begin{equation}
    \rho(x) := y(x) \sqrt{q(x)}
\end{equation}
and the new independent variable
\begin{equation}
    w(x) := \int q(x) dx + \mathcal{C}_w \label{w}
\end{equation}
where $q$ is an analytical function and $\mathcal{C}_w \in \mathbb{C}$. Rewriting our differential equation with respect to these variables yields \cite{wkb}
\begin{equation}
    \frac{d^2\rho}{dw^2} + (1 +\delta) \rho = 0
    \label{27}
\end{equation}
where we defined the function
\begin{equation}
    \delta(x) := -\frac{Q(x)}{q^2(x)\epsilon^2} - 1 + q^{-\frac{3}{2}}(x)\frac{d^2}{dx^2}\left(\frac{1}{\sqrt{q(x)}}\right).
    \label{68}
\end{equation}

The idea is to choose a function $q$ such that $|\delta| \sim 0$ in the region of the complex $x$-plane where we want to approximate the solution. Assuming we made such a choice, we can approximate our differential equation by
\begin{equation}
    \frac{d^2\rho}{dw^2} + \rho = 0
\end{equation}
whose general solution is
\begin{equation}
    \rho (w) = \mathcal{C}_+ e^{iw} + \mathcal{C}_- e^{-iw}.
    \label{28}
\end{equation}
for some $\mathcal{C}_\pm \in \mathbb{C}$. Now, we seek a function $q$ such that $|\delta| \sim 0$. If we choose
\begin{equation}
    q(x) = \frac{i\sqrt{Q(x)}}{\epsilon}
    \label{26}
\end{equation}
we can rewrite $\delta$ as
\begin{equation}
    \delta = \frac{\epsilon^2}{4}\left(\frac{1}{Q^2}\frac{d^2Q}{dx^2} -\frac{5}{4}\frac{1}{Q^3}\frac{dQ}{dx}\right).
    \label{188}
\end{equation}
Because $\epsilon^2$ is a small real parameter, the choice $(\ref{26})$ yields a small $\delta$ everywhere in the complex $x$-plane, except near roots of $Q$.

Plugging $(\ref{26})$ in $(\ref{28})$ yields the approximated solution

\begin{equation}
    y(x) = \frac{\mathcal{C}_-'}{\sqrt[4]{Q(x)}}\exp\left(-\frac{1}{\epsilon}\int\sqrt{Q(x)}dx\right) +  \frac{\mathcal{C}_+'}{\sqrt[4]{Q(x)}}\exp\left(\frac{1}{\epsilon}\int\sqrt{Q(x)}dx\right)
\end{equation}
for some $C_\pm'  \in \mathbb{C}$. We notice the solution above is identical to $(\ref{25})$!

Now, we want to know how good the expression above is, as an approximation to the true solution of $(\ref{24})$.

We already know the expression above does not accurately represent the solution near roots of $Q$. In the remaining zones of the complex $x$-plane, $\delta$ is small. We will now argue that this condition alone is not enough to ensure the expression above to be a good representation of the true solution. We start by expressing a true solution of the differential equation $(\ref{27})$ as
\begin{equation}
   \rho(w) = a_+(w)e^{iw} +  a_-(w)e^{-iw}
\end{equation}
for some analytical functions $a_+,a_-$. The expression above comes with an extra degree of freedom, for we are expressing the function $\rho$ using two arbitrary analytical functions $a_+,a_-$. To eliminate this freedom, we impose the condition
\begin{equation}
    \frac{da_+}{dw}e^{iw} +  \frac{da_-}{dw}e^{-iw} = 0.
\end{equation}
Using the condition above, we can write
\begin{equation}
    \frac{d\rho}{dw} = i a_+e^{iw} -i a_-e^{-iw}.
\end{equation}
The equations above allow us to rewrite the second order ordinary differential
equation $(\ref{27})$ as the system of two first order ordinary differential equations \cite{wkb}
\begin{equation}
    \frac{da_+}{dw}= \frac{i\delta}{2}\left(a_++a_-e^{-2iw}\right)
\end{equation}
\begin{equation}
    \frac{da_-}{dw}= -\frac{i\delta}{2}\left(a_-+a_+e^{2iw}\right).
    \label{207}
\end{equation}
Furthermore, we can rewrite the system above as the matrix differential equation
\begin{equation}
    \frac{da}{dw} = Ma
    \label{29}
\end{equation}
where we defined the matrices
\begin{equation}
    M(w) := \frac{i\delta}{2}\begin{bmatrix} 1 & e^{-2iw} \\  -e^{2iw} & -1
    \end{bmatrix} \hspace{20pt}  a(w) := \begin{bmatrix}
    a_+(w) \\ a_-(w)
    \end{bmatrix}.
\end{equation}

Looking at equations $(\ref{188})$ and $(\ref{29})$, we notice the relative change of $a$ with respect to $w$ is proportional to $\epsilon^2$. Thus, as long as the off diagonal elements of $M$ are of the order of unity, this change is negligible. In particular, we can ask for the condition
\begin{equation}
    \big\vert e^{-2iw}  \big\vert \sim  \big\vert  e^{2iw}  \big\vert \sim 1.
    \label{30}
\end{equation}

We found the validity domain we were looking for. Indeed, the solutions $(\ref{28})$ and consequently $(\ref{25})$ are good approximations of the respective true solutions, for values of $x$ where the condition above is fulfilled.

The condition above naturally leads to the definition of Stokes lines. Stokes lines are the curves in the complex $x$-plane where $w$ is real:
\begin{equation}
    \Im(w) = 0.
    \label{67}
\end{equation}

Naturally, from the definition (\ref{w}) the topology of these lines depends on the constant of integration $\mathcal{C}_w$. Moreover, along these lines $(\ref{30})$ holds true!

With this loose argument, we reason the solution $(\ref{25})$ is a good approximation of the true solution of $(\ref{24})$, along Stokes lines, sufficiently far away from roots of $Q$. A much more formal proof of this statement can be found in \cite{wkb}.

A final note should be taken regarding cases where an analytic continuation of $Q$ onto the complex $x$-plane cannot be made. As a general rule, we do not expect the solution $(\ref{25})$ to be a good approximation of the true solution, near singularities of $Q$. This is so, because $\delta$ might share the same singularities \cite{wkb}.

\section{A string-corrected black hole and its perturbations}
\label{cmp}
\noindent

In this section, the $d$-dimensional spherically symmetric black hole solution with leading string-theoretical higher derivative corrections obtained by Callan, Myers and Perry is introduced along with some results concerning gravitational and scalar field perturbations.

\subsection{Physical origin}
\noindent

String theories require corrections in powers of the inverse string tension $\alpha'$, which manifest themselves in the form of
higher-derivative terms in the effective action, and obviously in the corresponding field equations and their solutions.

We are focusing, in particular, in corrections to first order in $\alpha'$, which are present in bosonic and heterotic
string effective actions (but not on type II superstring).
The effective action we are thus considering is \cite{Callan}

\begin{equation} \label{eef}
\frac{1}{16 \pi G} \int \sqrt{-g} \left( \R -
\frac{4}{d-2} \left( \d^\mu \phi \right) \d_\mu \phi +
\mbox{e}^{\frac{4}{d-2} \phi} \frac{\lambda}{2}\
\R^{\mu\nu\rho\sigma} \R_{\mu\nu\rho\sigma} \right) \mbox{d}^dx
\end{equation}
where $\lambda = \frac{\alpha'}{2},\frac{\alpha'}{4},0$ for bosonic, heterotic and type II strings respectively. We are only considering gravitational
terms: we can consistently settle all fermions and gauge fields to zero. That is not the case of the dilaton, as it can be seen from the corresponding field equations.

The stringy correction present in the action above is bound to manifest itself as a correction term proportional to $\lambda$, and to affect measurable quantities associated with the black hole solution (e.g. the quasinormal modes we consider in this work). Thus, studying how these quantities change according to the stringy correction might be a viable way to provide experimental verifications of string theory.

\subsection{The metric}
\noindent

The Callan-Myers-Perry black hole we will be considering is a $d$-dimensional static spherically symmetric black hole of the form (\ref{35}). It is an asymptotically flat solution of the field equations from (\ref{eef}) - a $\lambda$-corrected Tangherlini black hole with $f(r)$ given by \cite{Callan}
\bea
f(r) := f_0(r)\left(1 + \lambda'\delta f\right), \label{fr2} \\
f_0(r) := 1- \left(\frac{R_h}{r}\right)^{d-3}, \\ \label{56}
\delta f (r) := -\frac{(d-3)(d-4)}{2}\left(\frac{R_h}{r}\right)^{d-3} \frac{1-\left(\frac{R_h}{r}\right)^{d-1}}{1-\left(\frac{R_h}{r}\right)^{d-3}}. \label{57}
\eea
In the equations above, $\lambda'$ stands for an adimensional version of $\lambda$, defined as $\lambda' := \frac{\lambda}{R_h^2}.$
Moreover, $R_h$ stands for the radius of the black hole event horizon, expressed as $R_h = \left(2\mu\right)^{\frac{1}{d-3}}$
where the parameter $\mu$ can be related to the black hole mass $m$ by
\begin{equation}
    m = \left(1 + \frac{(d-3)(d-4)}{2}\lambda'\right)\frac{(d-2)\Omega_{d-2}}{8 \pi G}\mu, \, \Omega_{d-2}=\frac{2 \pi^{\frac{d-1}{2}}}{\Gamma\left(\frac{d-1}{2}\right)}.
\end{equation}
We notice $R_h$ remained invariant under the stringy correction. This is a consequence of the coordinates used to express the metric $(\ref{35})$.

The Hawking temperature of this black hole is given, from (\ref{fr2}), by
\begin{equation}
    T_{\mathcal{H}} = \frac{1}{4\pi}\frac{df}{dr}(R_h) = \frac{d-3}{4\pi R_h}\left(1-\lambda'\frac{(d-1)(d-4)}{2}\right).
    \label{154}
\end{equation}

\subsection{Gravitational perturbations and quasinormal modes}
\noindent

Here, we introduce, very briefly, gravitational perturbations and present some associated results in the context of the Callan-Myers-Perry black hole (\ref{fr2}).

Linear perturbations of the metric are called gravitational perturbations. If we consider a spherically symmetric and static $d$-dimensional black hole space time of the form (\ref{35}), in the process of simplifying the equations of motion associated with the perturbation components one ends up with a simplified system of (decoupled) master equations \cite{tensortype}. The independent variables associated with these equations are functions that describe distinct types of gravitational perturbations. These types, scalar, vector and (for $d>4$) tensor types, are all needed in order to fully describe general gravitational perturbations. The master equations associated with these perturbations are second order Schr\"odinger-like differential equations of the form (\ref{36}), with $\psi_r$ replaced by $\psi$, a function describing the gravitational perturbations. In this context, $x$ is the tortoise coordinate corresponding to (\ref{fr2}).

This framework is valid in Einstein gravity and, as shown in different articles \cite{Takahashi:2009xh,Takahashi:2010ye,Dotti:2005sq,Gleiser:2005ra}, also in the presence of quadratic (Gauss-Bonnet) gravitational corrections.

In this work, we address gravitational perturbations associated with $d$-dimensional black holes with leading string gravitational corrections, like (\ref{eef}). We will be specifically concerned with tensor type gravitational perturbations of the black hole (\ref{fr2}), but our methods apply to other types of gravitational perturbations and other analogous black hole solutions \cite{Wheeler:1985nh,Moura:2009it}. In \cite{Moura:2006pz} and more generally in \cite{Moura:2012fq}, it was shown that the corresponding potential $V$ can be expressed, with respect to $r$, as
\begin{equation}
\begin{split}
    V[f(r)] = f(r)\left(\frac{l(l+d-3)}{r^2} + \frac{(d-2)(d-4)f(r)}{4r^2} + \frac{(d-2)f'(r)}{2r}\right) +\lambda' \\  \left(\frac{R_h}{r}\right)^2f(r)\Bigg[\left(\frac{2l(l+d-3)}{r} + \frac{(d-4)(d-5)f(r)}{r} + (d-4)f'(r)\right)\left(2\left(\frac{1-f(r)}{r}\right) + f'(r)\right) \\ +  \left(4(d-3)-(5d-16)f(r)\right)\frac{f'(r)}{r} - 4\left(f'(r)\right)^2+(d-4)f(r)f''(r)\Bigg]
    \label{37}
\end{split}
\end{equation}
where the apostrophe stands for derivation with respect to $r$.

We notice the potential above is identical to (\ref{153}) to zeroth order in $\lambda'$. Indeed, in the classical ($\lambda=0$) limit, linear perturbations of complex massless scalar fields and tensor type gravitational perturbations share the same quasinormal frequency spectrum \cite{Natario}.

\section{Quasinormal modes in the eikonal limit}
\label{qnmeik}
\noindent

In this section, we analytically compute the quasinormal frequencies, in the eikonal limit, of tensor type gravitational perturbations of the black hole (\ref{fr2}). For a general discussion see \cite{eikonal1,Cardoso:2008bp}.

We recall the eikonal limit targets quasinormal frequencies $\omega$ obeying the master equation (\ref{36}) for large values of the azimuthal number $l$.

\subsection{Definition and analysis of $Q$}
\noindent

From the potential (\ref{37}) we define the function
\begin{equation}
    Q(x) := \omega^2 - V(x).
\end{equation}
For large values of $l$, the asymptotic expansion of $Q$ can be written, with respect to $r$, as
\begin{equation}
    Q(r) \sim \omega^2 -h(r) \left(\frac{l}{r}\right)^2
    \label{38}
\end{equation}
where we defined
\begin{equation}
    h(r) := f(r)\left(1 + 2\lambda'\left(\frac{R_h}{r}\right)^2 \left[2(1-f(r))+rf'(r)\right]\right).
\end{equation}

Because we are working in the eikonal limit, we redefine $Q$ as the respective large $l$ asymptotic expansion $(\ref{38})$, for convenience purposes.

Up to zeroth order in $\lambda'$, we have
\begin{equation}
    Q(r) = \omega^2 -\left[1 - \left(\frac{R_h}{r}\right)^{d-3}\right] \left(\frac{l}{r}\right)^2.
\end{equation}
We notice the expression above only has one critical point, besides $r=0$, corresponding to one minimizer. Indeed, taking the derivative yields
\begin{equation}
    \frac{dQ}{dr}(r) =  l^2 \left(\frac{2}{r^3}-(d-1) R_h^{d-3} r^{-d}\right)
\end{equation}
Computing the root of the expression above, we get
\begin{equation}
    r_g := \left(\frac{d-1}{2}\right)^{\frac{1}{d-3}}R_h.
\end{equation}
Furthermore, taking the second order derivative yields
\begin{equation}
    \frac{d^2Q}{dr^2}(r_g) =   \frac{2^{\frac{d+1}{d-3}} (d-3) (d-1)^{-\frac{4}{d-3}} l^2}{R_h^4}.
    \label{41}
\end{equation}
Because the expression above is strictly positive for $d \ge 4$, we gather $r_g$ is a minimizer. When we consider $(\ref{38})$, up to first order in $\lambda'$, a correction of this minimizer emerges. To compute this correction, we redefine $r_g$ as the corrected minimizer
\begin{equation}
    r_g \mapsto r_g + \lambda' \delta r_g
\end{equation}
for some unknown $\delta r_g \in \mathbb{R}$ and replace it in the derivative of $Q$. The resulting expression, up to first order in $\lambda'$, is
\begin{equation}
\begin{split}
 \lambda' \frac{ 32^{\frac{1}{d-3}} (d-1)^{-\frac{2 (d+2)}{d-3}} l^2}{R_h^4}\Bigg(\delta r_g 2^{\frac{4-d}{3-d}} (d-3) (d-1)^{\frac{2 d}{d-3}}+ 4 ((d-11) d+16) (d-1)^{\frac{d+2}{d-3}} R_h+ \\ \left(4 (d+1)-4^{\frac{1}{3-d}} (d-4) (d-3) (d-1)^{\frac{2}{d-3}}\right) (d-1)^{\frac{2d-1}{d-3}} R_h\Bigg).
  \end{split}
\end{equation}
In order to find $\delta r_g$, we equate the expression above to zero and solve for $\delta r_g$. After some algebraic manipulation, we get
\begin{equation}
  \delta r_g =  \left(\frac{d-1}{2}\right)^{\frac{1}{d-3}}\left(\frac{d-4}{2}-(2d-5)\left(\frac{2}{d-1}\right)^{\frac{d-1}{d-3}}\right)R_h.
\end{equation}

Finally, we can write the corrected minimizer as
\begin{equation}
    r_g = \left(\frac{d-1}{2}\right)^{\frac{1}{d-3}}R_h\left[1 +\lambda'\left(\frac{d-4}{2}-(2d-5)\left(\frac{2}{d-1}\right)^{\frac{d-1}{d-3}}\right) \right].
    \label{200}
\end{equation}

\subsection{WKB approximation}
\noindent

\begin{figure}[h]
\centering
\includegraphics[width=0.5\textwidth]{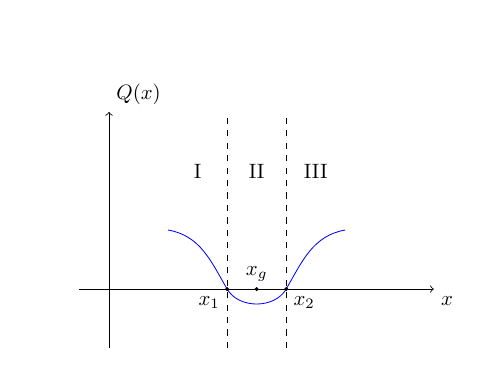}
\caption{Graphical depiction of $Q$ as the blue line. Possible roots of $Q$ are denoted by $x_1$ and $x_2$. Moreover, the x-axis is subdivided in three regions, denoted by I, II and III.}
\label{fig4}
\end{figure}

We now use the WKB theory to find an analytical relation between quasinormal frequencies, in the eikonal limit, and the potential $(\ref{37})$. To this end, we resort to the previously found minimum of $Q$. We start by making the assumption that quasinormal modes $\omega$, in the eikonal limit, are such that
\begin{equation}
    \omega^2 \sim V(r_g).
    \label{39}
\end{equation}
If this is the case, $Q$ can be graphically represented near $x_g = x(r_g)$, as depicted in figure \ref{fig4}.

 \noindent

In this figure, we notice the x-axis is subdivided in three regions, denoted by I, II and III.

Looking at $(\ref{38})$, we notice $|Q|$ is very large far enough from $x_g$ and decreases abruptly, approaching the latter. Thus, we can use WKB theory to approximate the solution of the master equation $(\ref{36})$ in regions I and III. Using $(\ref{25})$ yields
\begin{equation}
    \psi_{\text{I}}(x) \sim \frac{\mathcal{C}_1^+}{\sqrt[4]{Q(x)}}\exp\left( i \int_{x}^{x_1}\sqrt{Q(y)}dy\right) + \frac{\mathcal{C}_1^-}{\sqrt[4]{Q(x)}}\exp\left( -i \int_{x}^{x_1}\sqrt{Q(y)}dy\right)
\end{equation}
\begin{equation}
    \psi_{\text{III}}(x) \sim \frac{\mathcal{C}_2^+}{\sqrt[4]{Q(x)}}\exp\left( i \int_{x_2}^{x}\sqrt{Q(y)}dy\right) + \frac{\mathcal{C}_2^-}{\sqrt[4]{Q(x)}}\exp\left( -i \int_{x_2}^{x}\sqrt{Q(y)}dy\right)
\end{equation}
for some $\mathcal{C}_1^\pm,\mathcal{C}_2^\pm \in \mathbb{C}$. To compute an approximation for $\psi_{\text{II}}$, we start by considering the Taylor expansion
\begin{equation}
    Q(x) = Q(x_g) + \frac{dQ}{dx}(x_g)(x-x_g) + \frac{d^2Q}{dx^2}(x_g)\frac{(x-x_g)^2}{2} + \mathcal{O}\left(\left(x-x_g\right)^3\right).
    \label{40}
\end{equation}
Noting that
\begin{equation}
    \frac{dQ}{dx}(x_g) = 0
\end{equation}
we can rewrite $(\ref{40})$ as
\begin{equation}
    Q(x) = Q(x_g)  + \frac{d^2Q}{dx^2}(x_g)\frac{(x-x_g)^2}{2} + \mathcal{O}\left(\left(x-x_g\right)^3\right).
\end{equation}
Furthermore, we can neglect terms of higher orders in $(x-x_g)$ and write
\begin{equation}
    Q(x) \sim Q(x_g)+ \frac{d^2Q}{dx^2}(x_g)\frac{(x-x_g)^2}{2}.
    \label{42}
\end{equation}
in region II. This approximation is valid because region II is very slim, under the assumption $(\ref{39})$. Now, defining the constants
\begin{equation}
    k := \frac{1}{2}\frac{d^2Q}{dx^2}(x_g) \hspace{20pt}  \nu := - \frac{1}{2} -iQ(x_g)\left(2\frac{d^2Q}{dx^2}(x_g)\right)^{-\frac{1}{2}}
    \label{189}
\end{equation}
and the variable
\begin{equation}
    t := \sqrt[4]{4k}e^{\frac{i\pi}{4}}(x-x_g)
\end{equation}
allow us to rewrite the master equation $(\ref{36})$ as
\begin{equation}
    \frac{d^2 \psi_{\text{II}}}{dt^2} + \left(\nu + \frac{1}{2} -\frac{t^2}{4}\right)\psi_{\text{II}}=0
\end{equation}
in region II. The general solution of the differential equation above is
\begin{equation}
    \psi_{\text{II}}(t) = \mathcal{C}_1D_\nu(t) + \mathcal{C}_2D_{-\nu-1}(it)
\end{equation}
for some $\mathcal{C}_1,\mathcal{C}_2 \in \mathbb{C}$, where
\begin{equation}
    D_\nu(w) = \frac{2^{\frac{\nu}{2}}e^{-\frac{w^2}{4}}\sqrt[4]{-iw}\sqrt[4]{iw}}{\sqrt{w}}\,_1F_1\left(-\frac{\nu}{2},\frac{1}{2},\frac{w^2}{2}\right)
\end{equation}
denotes the parabolic cylinder function and
\begin{equation}
    \,_1F_1\left(a,b,w\right) = \sum_{k=0}^{+\infty }\frac{\Gamma\left(a+k\right)}{\Gamma(a)}\frac{\Gamma(b)}{\Gamma(b+k)}\frac{w^k}{k!}
\end{equation}
 stands for the confluent hypergeometric function of the first kind.

 The asymptotic expansions of the parabolic cylinder function yield \cite{eikonal1}
 \begin{equation}
 \begin{split}
     \psi_{\text{II}}(x) \sim \mathcal{C}_2e^{-3i\pi\frac{\nu+1}{4}}(4k)^{-\frac{\nu+1}{4}}(x-x_g)^{-(\nu+1)}e^{i\sqrt{k}\frac{(x-x_g)^2}{2}}  \\ +\left(\mathcal{C}_1 + \mathcal{C}_2\frac{\sqrt{2\pi}e^{-i\nu\frac{\pi}{2}}}{\Gamma(\nu+1)}\right)e^{i\pi\frac{\nu}{4}}
     (4k)^{\frac{\nu}{4}}\left(x-x_g\right)^{\nu}e^{-i\sqrt{k}\frac{(x-x_g)^2}{2}}
     \label{47}
     \end{split}
 \end{equation}
 \begin{equation}
 \begin{split}
     \psi_{\text{II}}(x) \sim \mathcal{C}_1e^{-3i\pi\frac{\nu}{4}}(4k)^{\frac{\nu}{4}}(x_g-x)^{\nu}e^{-i\sqrt{k}\frac{(x-x_g)^2}{2}}  \\ +\left(\mathcal{C}_2 -i \mathcal{C}_1\frac{\sqrt{2\pi}e^{-i\nu\frac{\pi}{2}}}{\Gamma(-\nu)}\right)e^{i\pi\frac{\nu+1}{4}}
     (4k)^{-\frac{\nu+1}{4}}\left(x_g-x\right)^{-(\nu+1)}e^{i\sqrt{k}\frac{(x-x_g)^2}{2}} \label{48}
     \end{split}
 \end{equation}
 for $te^{-i\frac{\pi}{4}} \gg 1$ and $te^{-i\frac{\pi}{4}} \ll -1$ respectively. Looking at $(\ref{41})$, we see that $k \gg 1$. Hence, the asymptotic expansions above are valid for values of $x$ near the respective points $x_1$ and $x_2$, where it still makes sense to match them with $\psi_{\text{I}}$ and $\psi_{\text{III}}$.

 Imposing the boundary conditions $(\ref{20})$ and $(\ref{21})$ on solutions $\psi_{\text{I}}$ and $\psi_{\text{III}}$ respectively yields
 \begin{equation}
     \mathcal{C}_1^+ = 0   \hspace{20pt} \mathcal{C}_2^+ = 0.
 \end{equation}
Thus, we can rewrite $\psi_{\text{I}}$ and $\psi_{\text{III}}$ as
 \begin{equation}
     \psi_{\text{I}}(x) \sim \frac{\mathcal{C}_1^-}{\sqrt[4]{Q(x)}}\exp\left(-i\int_{x}^{x_1}\sqrt{Q(y)}dy\right)
 \end{equation}
  \begin{equation}
     \psi_{\text{III}}(x) \sim \frac{\mathcal{C}_2^-}{\sqrt[4]{Q(x)}}\exp\left(-i\int_{x_2}^{x}\sqrt{Q(y)}dy\right).
 \end{equation}
 In the matching region, we still can use the approximation $(\ref{42})$ and write
 \begin{equation}
      \psi_{\text{I}}(x) \sim \frac{\mathcal{C}_1^-}{\sqrt[4]{Q(x)}}\exp\left(-i\int_{x}^{x_1}\sqrt{Q(x_g) +k(y-x_g)^2}dy\right)
      \label{43}
 \end{equation}
  \begin{equation}
      \psi_{\text{III}}(x) \sim \frac{\mathcal{C}_2^-}{\sqrt[4]{Q(x)}}\exp\left(-i\int_{x_2}^{x}\sqrt{Q(x_g) +k(y-x_g)^2}dy\right).
      \label{44}
 \end{equation}
Under the assumption $(\ref{39})$, we know that
\begin{equation}
    Q(x_g) \sim 0.
\end{equation}
Thus, we can write
\begin{equation}
\begin{split}
   \int_{x}^{x_1}\sqrt{Q(x_g) + k(y-x_g)^2}dy \sim \int_{x}^{x_1}\sqrt{k(y-x_g)^2}dy = \sqrt{k}\int_{x}^{x_1}\big\vert y-x_g\big\vert dy \\= -\sqrt{k} \int_{x}^{x_1}(y-x_g)dy = \sqrt{k}\frac{(x-x_g)^2}{2} - \sqrt{k}\frac{(x_1-x_g)}{2},
    \end{split}
\end{equation}
\begin{equation}
\begin{split}
    \int_{x_2}^x\sqrt{Q(x_g) + k(y-x_g)^2}dy \sim\int_{x_2}^{x}\sqrt{k(y-x_g)^2}dy = \sqrt{k}\int_{x_2}^{x}\big\vert y-x_g\big\vert dy = \\ \sqrt{k} \int_{x_2}^{x}(y-x_g)dy = \sqrt{k}\frac{(x-x_g)^2}{2} - \sqrt{k}\frac{(x_2-x_g)}{2}
    \end{split}
\end{equation}
Using the approximations above, we can rewrite $(\ref{43})$ and $(\ref{44})$ as
\begin{equation}
     \psi_{\text{I}}(x) \sim \frac{\mathcal{D}_1}{\sqrt[4]{Q(x)}}e^{-i\sqrt{k}\frac{(x-x_g)^2}{2}},
     \label{45}
\end{equation}
\begin{equation}
     \psi_{\text{III}}(x) \sim \frac{\mathcal{D}_2}{\sqrt[4]{Q(x)}}e^{-i\sqrt{k}\frac{(x-x_g)^2}{2}}.
     \label{46}
\end{equation}
for some $\mathcal{D}_1,\mathcal{D}_2 \in \mathbb{C}$. We notice the exponential terms of the asymptotic expansions $(\ref{47})$ and $(\ref{48})$ match those of $(\ref{46})$ and $(\ref{45})$ respectively if and only if $\mathcal{C}_2=0$ and
\begin{equation}
    \frac{1}{\Gamma(-\nu)}\sim 0.
\end{equation}
The last condition demands $\nu$ to be very close to singular points of $\Gamma$. Thus, we end up with the restriction
\begin{equation}
    \nu \in \mathbb{N}_0.
\end{equation}
Using definition $(\ref{189})$, we can rewrite the condition above as
\begin{equation}
    Q(x_g)\left(2\frac{d^2Q}{dx^2}(x_g)\right)^{-\frac{1}{2}} = i \left(n + \frac{1}{2}\right)
    \label{49}
\end{equation}
for $n \in \mathbb{N}_0$. This is the condition we use to analytically compute the quasinormal frequencies.

\subsection{Quasinormal frequencies computation}
\noindent

We now explicitly compute an analytical expression for the quasinormal frequencies in the eikonal limit. Looking at the asymptotic expression $(\ref{38})$, we easily see that
\begin{equation}
    2h(r_g) = r_g \frac{dh}{dr}(r_g).
    \label{51}
\end{equation}
Using $(\ref{38})$, we can rewrite the equation $(\ref{49})$ as
\begin{equation}
    \left(\omega^2-h(r_g)\left(\frac{l}{r_g}\right)^2\right) = i\left(n +\frac{1}{2}\right)\sqrt{2\frac{d^2Q}{dx^2}(x_g)}
\end{equation}
for $n \in \mathbb{N}_0$. Rearranging some terms yields
\begin{equation}
    \omega = \pm l\sqrt{\frac{h(r_g)}{r_g^2}} \sqrt{1 + i \left(n+\frac{1}{2}\right)\frac{r_g^2}{l^2h(r_g)}\sqrt{2\frac{d^2Q}{dx^2}(x_g)}}.
    \label{52}
\end{equation}
Once more, using $(\ref{38})$ yields
\begin{equation}
\begin{split}
    \frac{d^2Q}{dx^2}(x_g) = -l^2 \frac{d^2}{dx^2}\left(\frac{h}{r^2}\right)(x_g) = - l^2\frac{d}{dx}\left[\left(\frac{dh}{dr}\frac{1}{r^2} - \frac{2}{r^3}h\right)\frac{dr}{dx}\right](x_g) \\ = -l^2 \frac{dr}{dx}\left[ \frac{d^2r}{drdx}\left(\frac{dh}{dr}\frac{1}{r^2}-\frac{2}{r^3}h\right) +\frac{dr}{dx}\left(\frac{d^2h}{dr^2}\frac{1}{r^2} - \frac{2}{r^3}\frac{dh}{dr} + \frac{6}{r^4}h - \frac{2}{r^3}\frac{dh}{dr}\right) \right](x_g).
    \end{split}
\end{equation}
Using definition $(\ref{50})$ and equation $(\ref{51})$, we can rewrite the previous equation as
\begin{equation}
    \frac{d^2Q}{dx^2}(x_g) = -l^2\frac{f^2(r_g)}{r_g^4}\left(\frac{d^2h}{dr^2}(r_g)r_g^2 - 2h(r_g)\right).
\end{equation}
Hence, we can rewrite equation $(\ref{52})$ as
\begin{equation}
    \omega = \pm l\sqrt{\frac{h(r_g)}{r_g^2}}\sqrt{1+\frac{i}{l}\left(n+\frac{1}{2}\right)\sqrt{2\left(\frac{f(r_g)}{h(r_g)}\right)^2\left(2h(r_g)-\frac{d^2h}{dr^2}(r_g)r_g^2\right)}}
\end{equation}
for $n \in \mathbb{N}_0$. Since we are in the eikonal limit, we can Taylor expand the expression above yielding
\begin{equation}
    \omega =  l\sqrt{\frac{h(r_g)}{r_g^2}}  +\frac{i}{\sqrt{2}}\left(n+\frac{1}{2}\right)\sqrt{\frac{f^2(r_g)}{r_g^2h(r_g)}\left(2h(r_g)-\frac{d^2h}{dr^2}(r_g)r_g^2\right)}
    \label{59}
\end{equation}
 where we chose the positive root, assuming stability of the black hole solution. In order to simplify the notation, we define the constants
\begin{equation}
    \Gamma_r := \sqrt{\frac{h(r_g)}{r_g^2}}, \hspace{20pt} \Gamma_i := \sqrt{\frac{f^2(r_g)}{2r_g^2h(r_g)}\left(2h(r_g)-\frac{d^2h}{dr^2}(r_g)r_g^2\right)}.
\end{equation}
After some algebraic manipulation, we can write $\Gamma_r$ and $\Gamma_i$, up to first order in $\lambda'$, as
\begin{equation}
    \Gamma_r = \sqrt{\frac{d-3}{d-1}}\left(\frac{2}{d-1}\right)^{\frac{1}{d-3}}\frac{1}{R_h}\left[1+\frac{\lambda'}{2}\left(3(d-2)\left(\frac{2}{d-1}\right)^{\frac{d-1}{d-3}} -(d-4)\right)\right],
\end{equation}
\begin{equation}
    \Gamma_i = \frac{d-3}{\sqrt{d-1}} \left(\frac{2}{d - 1}\right)^{\frac{1}{d-3}} \frac{1}{R_h} \left[1 -\lambda'\frac{d-4}{2} \left(
1+ (d-2) \left(\frac{2}{d - 1}\right)^{\frac{d-1}{d-3}} \right) \right].
\label{192}
\end{equation}
Rewriting the expressions above with respect to the black hole Hawking temperature $T_{\mathcal{H}}$, given by (\ref{154}), yields
\begin{equation}
    \Gamma_r =  \frac{4 \pi T_{\mathcal{H}}}{\sqrt{(d-1)(d-3)}} \left(\frac{2}{d - 1}\right)^{\frac{1}{d-3}} \left[1 +  \lambda'\left(\frac{d-2}{2}\right) \left(d-4+3\left(\frac{2}{d-1}\right)^{\frac{d-1}{d-3}} \right) \right],
\end{equation}
\begin{equation}
     \Gamma_i = \frac{4\pi T_{\mathcal{H}}}{\sqrt{d-1}}\left(\frac{2}{d-1}\right)^{\frac{1}{d-3}} \left[1 +  \lambda' \frac{(d-2)(d-4)}{2} \left( 1-\left(\frac{2}{d - 1}\right)^{\frac{d-1}{d-3}} \right) \right].
     \label{210}
\end{equation}
Finally, we can rewrite the quasinormal frequencies $(\ref{59})$ as
\begin{equation}
    \omega = l\Gamma_r + i\left(n + \frac{1}{2}\right)\Gamma_i
    \label{209}
\end{equation}
for $n \in \mathbb{N}_0$. We notice they obey the starting assumption $(\ref{39})$.

This concludes our derivation of the quasinornal frequencies for tensor type gravitational perturbations of the black hole (\ref{fr2}) in the eikonal limit. This derivation can also be found in \cite{Moura:2021eln}, together with an analogous calculation for test scalar fields in the same background. Quasinornal frequencies in the eikonal limit for generic gravitational perturbations of black holes with Gauss Bonnet corrections can be found in \cite{Konoplya:2017wot}.

\section{Quasinormal modes in the asymptotic limit}
\label{qnmd}
\noindent

In this section, we analytically compute the quasinormal frequencies, in the asymptotic limit, of tensor type gravitational perturbations of the black hole (\ref{fr2}). We recall the asymptotic limit targets quasinormal frequencies $\omega$ such that $|\Im(\omega)|\gg |\Re(\omega)|$. The corresponding quasinormal modes are therefore highly damped.

First, we address the master equation $(\ref{36})$ by making a crucial variable change followed by a standard perturbation theory approach.

Secondly, we set up some of the WKB theory formalism and study the associated Stokes lines topology.

Finally, we introduce the monodromy method to analytically compute the quasinormal frequencies. The monodromy method, introduced in \cite{motl}, is a very powerful method allowing one to analytically compute quasinormal frequencies, in the asymptotic limit, associated with linear perturbations of fields coupled to gravity in a black hole space time \cite{Natario}.

In this section we allow $r$ to take complex values. Thus, we assume an analytical continuation of functions of $r$ to the complex plane.

\subsection{Variable change and perturbation theory}
\noindent

For future convenience, we want to rewrite the master equation $(\ref{36})$ with respect to a simpler independent variable. The variable we seek is the tortoise coordinate of the $d$-dimensional Tangherlini black hole space time. Recalling (\ref{56}), this variable is defined as
\begin{equation}
    dz = \frac{dr}{f_0}
    \label{62}
\end{equation}
up to a complex integration constant. We allow the integration constant to be complex, because we are assuming an analytical continuation of the functions of $r$ to the complex plane.

Looking at definitions $(\ref{50})$ and $(\ref{62})$, we can write
\begin{equation}
    \frac{dz}{dx} = 1+\lambda'\delta f.
    \label{63}
\end{equation}
We can also rewrite the second order derivative of the master equation $(\ref{36})$ as
\begin{equation}
    \frac{d^2\psi}{dx^2} = \frac{d}{dx}\left(\frac{d\psi}{dz}\frac{dz}{dx}\right) = \frac{d^2\psi}{dz^2}\left(\frac{dz}{dx}\right)^2 + \frac{d\psi}{dz}\frac{d^2z}{dx^2}  = \frac{d^2\psi}{dz^2}\left(\frac{dz}{dx}\right)^2 + \frac{d}{dr}\left(\frac{dz}{dx}\right)\frac{dr}{dx}\frac{d\psi}{dz}.
\end{equation}
Thus, we can rewrite the master equation $(\ref{36})$, with respect to $z$, as
\begin{equation}
    \left(\frac{dz}{dx}\right)^2\frac{d^2\psi}{dz^2} + \frac{d}{dr}\left(\frac{dz}{dx}\right)\frac{dr}{dx}\frac{d\psi}{dz} + \left(\omega^2 - V\right)\psi = 0.
    \label{64}
\end{equation}
Using $(\ref{50})$ and $(\ref{63})$, we can write, with respect to $r$, the expressions
\begin{equation}
   \frac{d}{dr}\left(\frac{dz}{dx}\right)\frac{dr}{dx}(r)=\lambda'\frac{(d-4) (d-3)  r^{2-2 d} \left((d-1) r^4 R_h^{2 d} -2 (d-2) R_h^{d+3} r^{d+1} +(d-3) R_h^4 r^{2 d}\right)}{2 R_h^{7-d} r^d-2 R_h^4 r^3}
\end{equation}
\begin{equation}
    \left(\frac{dz}{dx}\right)^2(r) = 1 + 2\lambda'\delta f = 1 - \lambda'(d-3)(d-4)\left(\frac{R_h}{r}\right)^{d-3}\frac{1-\left(\frac{R_h}{r}\right)^{d-1}}{1-\left(\frac{R_h}{r}\right)^{d-3}}
\end{equation}
up to first order in $\lambda'$.

Now, we use perturbation theory to address the differential equation $(\ref{64})$. We start by considering the expansions
\begin{equation}
    \psi = \psi_0 + \lambda' \psi_1
    \label{99}
\end{equation}
\begin{equation}
    V = V_0 + \lambda'V_1.
\end{equation}

Here, $V_0$ stands for the potential $(\ref{37})$ in the limit of $\lambda'$ set to zero, i.e. the potential (\ref{153}) taken with $f_0(r)$. Moreover, $V_1$ contains all the corrections of first order in $\lambda'$. This includes those coming from the explicit correction present in (\ref{37}) and those coming from the correction present in $f$ in (\ref{fr2}).

Replacing these expansions in (\ref{64}) and solving perturbatively in powers of $\lambda'$ yields two distinct linear ordinary differential equations. The first one, of zeroth order in $\lambda'$, is
\begin{equation}
    \frac{d^2\psi_0}{dz^2} + \left(\omega^2 -V_0\right)\psi_0 = 0.
    \label{65}
\end{equation}
The second one, of first order in $\lambda'$, is the non homogeneous differential equation
\begin{equation}
    \frac{d^2\psi_1}{dz^2} + \left(\omega^2-V_0\right)\psi_1 = \xi.
    \label{66}
\end{equation}
The non homogeneous term is
\begin{equation}
    \xi = \xi_1 \frac{d^2\psi_0}{dz^2} + \xi_2 \frac{d\psi_0}{dz} + \xi_3\psi_0
    \label{85}
\end{equation}
where we defined the functions
\begin{equation}
    \xi_1(r) := -2\delta f(r)
    \label{77}
\end{equation}
\begin{equation}
    \xi_2(r):= - f(r)\left[\delta f\right]'(r)
    \label{78}
\end{equation}
\begin{equation}
    \xi_3(r):= V_1(r)
    \label{79}
\end{equation}

From now on, we restrict ourselves to the asymptotic limit. Hence, we assume that $|\Im\left(\omega\right)| \gg |\Re\left(\omega\right)|$. In the asymptotic limit, some of the WKB theory formalism developed in section \ref{wkb} can be used to give a viable approximation of the general solutions of the differential equations $(\ref{65})$ and $(\ref{66})$. In the next section, we set up this formalism and proceed to compute this approximation for the general solution of $(\ref{65})$.

\subsection{WKB theory set up}
\noindent

We now set up the WKB theory formalism associated with the differential equation (\ref{65}).

We note it is preferable to rewrite (\ref{65}) with respect to the coordinate $r$, instead of $z$. This is so, mainly because, as we will see in (\ref{69}), $z$ is a multivalued function of $r$. Following \cite{andersson}, we define a new dependent variable
\begin{equation}
    \Psi := \sqrt{f_0}\psi_0.
    \label{53}
\end{equation}
Now, we can write
\begin{equation}
    \frac{d^2\psi_0}{dz^2} = \frac{d}{dz}\left(\frac{d\psi_0}{dr}\frac{dr}{dz}\right) = \frac{d^2\psi_0}{dr^2}\left(\frac{dr}{dz}\right)^2 + \frac{d\psi_0}{dr} \frac{d^2r}{drdz}\frac{dr}{dz}.
\end{equation}
Using definition $(\ref{62})$ allows us to rewrite the previous equation as
\begin{equation}
    \frac{d^2\psi_0}{dx^2} = \frac{d^2\psi_0}{dr^2}f_0^2+f_0'f_0\frac{d\psi_0}{dr}
    \label{54}
\end{equation}
where the apostrophe stands for derivation with respect to $r$. Using definition $(\ref{53})$, we can write
\begin{equation}
\frac{d\psi_0}{dr} = -\frac{1}{2}f_0^{-\frac{3}{2}}f_0'\Psi +\frac{1}{\sqrt{f_0}}\frac{d\Psi}{dr}
\end{equation}
\begin{equation}
    \frac{d^2\psi_0}{dr^2} = \frac{3}{4}f_0^{-\frac{5}{2}}\left(f_0'\right)^2\Psi -\frac{1}{2}f_0^{-\frac{3}{2}}f_0''\Psi -f_0^{-\frac{
    3}{2}}f_0'\frac{d\Psi}{dr} + \frac{1}{\sqrt{f_0}}\frac{d^2\Psi }{dr^2}.
\end{equation}
Replacing these expressions in equation $(\ref{54})$ allow us to rewrite $(\ref{65})$ as
\begin{equation}
    \frac{d^2\Psi}{dr^2} + R\Psi = 0.
    \label{55}
\end{equation}
where we defined the function
\begin{equation}
    R := \left(\frac{\omega}{f_0}\right)^2 - \frac{V_0}{f^2_0} - \frac{1}{2}\frac{f_0''}{f_0} + \frac{1}{4}\left(\frac{f_0'}{f_0}\right)^2.
    \label{58}
\end{equation}

In the asymptotic limit, quasinormal frequencies are such that $|\omega| \gg 1$. Hence, we can use $(\ref{25})$ to write the WKB approximation of the general solution of $(\ref{55})$ for some $\mathcal{C}_\pm\in \mathbb{C}$ as
\begin{equation}
    \Psi(r) \sim \mathcal{C}_+\sqrt[4]{\frac{\omega^2}{R(r)}}\exp\left(i\omega\int\sqrt{\frac{R(r)}{\omega^2}}dr\right) + \mathcal{C}_-\sqrt[4]{\frac{\omega^2}{R(r)}}\exp\left(-i\omega\int\sqrt{\frac{R(r)}{\omega^2}}dr\right).
    \label{60}
\end{equation}

Looking at $(\ref{58})$, we notice $R$ has singularities in roots of $f_0$ and in $r=0$. Considering the definition $(\ref{56})$, we figure these roots are solutions of the polynomial equation
\begin{equation}
    1 - \left(\frac{R_h}{r}\right)^{d-3} = 0
\end{equation}
such as the event horizon $r=R_h$. The remaining $d$-4 solutions, usually called ``fictitious horizons'', can be found equally distributed among the circle $|r| = R_h$. Furthermore, because we are considering the asymptotic limit and consequently assuming that $|\omega|\gg 1$, we know the roots of $R$ are very close to $r = 0$. Thus, we expect $(\ref{60})$ to provide a good approximation of the general solution of the differential equation $(\ref{55})$, in Stokes lines, far enough from $r=0$ and from the real and fictitious horizons. In these regions, the approximation
\begin{equation}
     R \sim \left(\frac{\omega}{f_0}\right)^2
    \label{190}
\end{equation}
holds true. Hence, we can rewrite the WKB approximation $(\ref{60})$ as
\begin{equation}
    \Psi(r) \sim \mathcal{C}_+\sqrt{f_0(r)}\exp\left(i\omega\int\frac{dr}{f_0(r)}\right) +\mathcal{C}_-\sqrt{f_0(r)}\exp\left(-i\omega\int\frac{dr}{f_0(r)}\right)
\end{equation}
in these regions. Using definition $(\ref{62})$, we can rewrite the approximation above as
\begin{equation}
    \Psi(r) \sim \mathcal{C}_+\sqrt{f_0(r)}e^{i\omega z(r)} + \mathcal{C}_-\sqrt{f_0(r)}e^{-i\omega z(r)}.
\end{equation}
Finally, using definition $(\ref{53})$ and omitting the $r$ dependence for simplicity purposes, we can rewrite the approximation above as
\begin{equation}
    \psi_0(z) \sim \mathcal{C}_+e^{i\omega z} + \mathcal{C}_-e^{-i\omega z}.
    \label{61}
\end{equation}
We will use this approximation several times during this work.

\subsection{Stokes lines topology}
\noindent

Now, we want to know the regions of the complex $r$-plane where the WKB approximation $(\ref{61})$ is effective. As we saw in section \ref{wkb}, these regions will generally coincide with Stokes lines associated with the approximation. Therefore, in order to know these regions well, we need to start by studying the Stokes lines topology associated with the WKB approximation $(\ref{61})$.

Considering the approximation $(\ref{190})$ and definition $(\ref{67})$, Stokes lines, far enough from $r = 0$ and from the real and fictitious horizons, are such that
\begin{equation}
    \Im\left(\omega z\right) =0 \Rightarrow \Re\left(z\right) =0.
\end{equation}
In the first equality above, we used the fact that we are in the asymptotic limit and consequently assuming $\omega$ to be approximately imaginary.

It is rather easy to see the topology of these lines in the surroundings of $r=0$. Indeed, integrating (\ref{62}) yields \cite{Natario}
\begin{equation}
    z(r) = r + \frac{1}{2}\sum_{n=0}^{d-4}\frac{1}{k_n}\log\left(1-\frac{r}{R_n}\right)
    \label{69}
\end{equation}
where we defined
\begin{equation}
    k_n := \frac{1}{2}f_0'(R_n)
    \label{70}
\end{equation}
and where we chose the complex integration constant, defining $z$, to be such that $z(r=0) = 0$. In the expressions above, $R_n$ denotes fictitious horizons for $1 \le n \le d-4$ and the real horizon for $n = 0$.

For values of $r$ sufficiently close to the origin of the complex $r$-plane, the logarithm in $(\ref{69})$ is given by the Taylor series
\begin{equation}
    \log\left(1 - \frac{r}{R_n}\right) = -\sum_{k=1}^{+\infty} \frac{1}{k}\left(\frac{r}{R_n}\right)^k.
\end{equation}
The lowest order contribution from the Taylor series above to the sum in $(\ref{69})$ is
\begin{equation}
     -\frac{1}{2}\sum_{n=0}^{d-4}\frac{r}{k_nR_n}.
\end{equation}
Using definitions $(\ref{56})$ and $(\ref{70})$, we can rewrite the sum above as
\begin{equation}
    -\frac{1}{2}\sum_{n=0}^{d-4}\frac{r}{k_nR_n} = -\sum_ {n=0}^{d-4}\frac{r}{(d-3)} = -r.
\end{equation}
Thus, we notice the leading order contribution from the truncated Taylor expansion of the logarithm is cancelled by the first term in $(\ref{69})$. As a consequence, the dominant term of $z$, near the origin of the complex $r$-plane, is provided by the next nonzero contribution from the Taylor expansion of $(\ref{69})$. It is easy to see that
\begin{equation}
    \sum_{n=0}^{d-4} \frac{1}{k_n}\left(\frac{r}{R_n}\right)^k = \sum_ {n=0}^{d-4} \frac{R_n^{d-2}}{(d-3)R_h^{d-3}}\left(\frac{r}{R_n}\right)^k = \frac{r^k}{d-3}\sum_ {n=0}^{d-4} R_n^{1-k} = 0
\end{equation}
for $1 < k <d-2$. Indeed, noting the horizons are equally distributed along the circle $|r| = R_h$, the last equality above follows easily. For $k = d-2$, we have the contribution
\begin{equation}
    -\frac{1}{d-2}\sum_{n=0}^{d-4}\frac{r^{d-2}}{d-3}R_h^{3-d} =  - \frac{R_h^{3-d}}{d-2}\sum_ {n=0}^{d-4}\frac{r^{d-2}}{d-3} = - \frac{R_h^{3-d}r^{d-2}}{d-2}.
\end{equation}
Thus, we can write
\begin{equation}
    z(r) \sim - \frac{R_h^{3-d}r^{d-2}}{d-2}.
    \label{76}
\end{equation}
near the surroundings of $r=0$.

\begin{figure}[h]
\begin{subfigure}{0.32\textwidth}
\includegraphics[width=5.2cm, height=5.2cm]{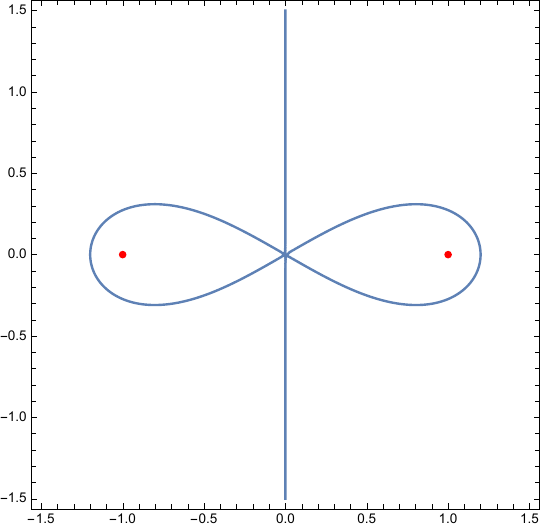}
\caption{$d=5$}
\end{subfigure}
\begin{subfigure}{0.32\textwidth}
\includegraphics[width=5.2cm, height=5.2cm]{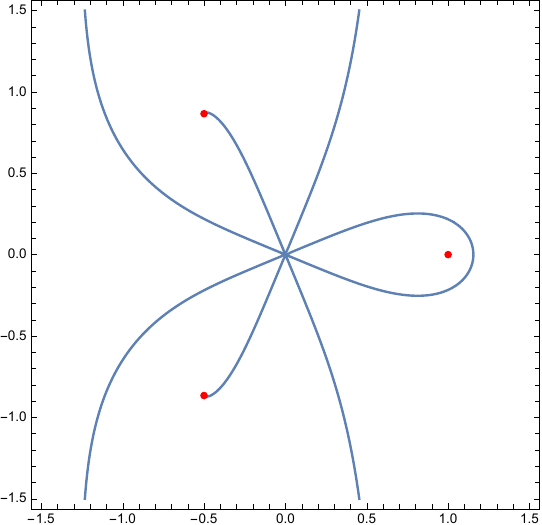}
\caption{$d=6$}
\end{subfigure}
\begin{subfigure}{0.32\textwidth}
\includegraphics[width=5.2cm, height=5.2cm]{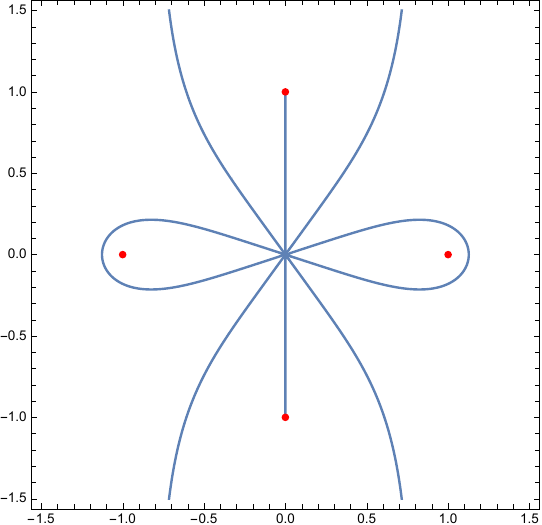}
\caption{$d=7$}
\end{subfigure}
\begin{subfigure}{0.32\textwidth}
\includegraphics[width=5.2cm, height=5.2cm]{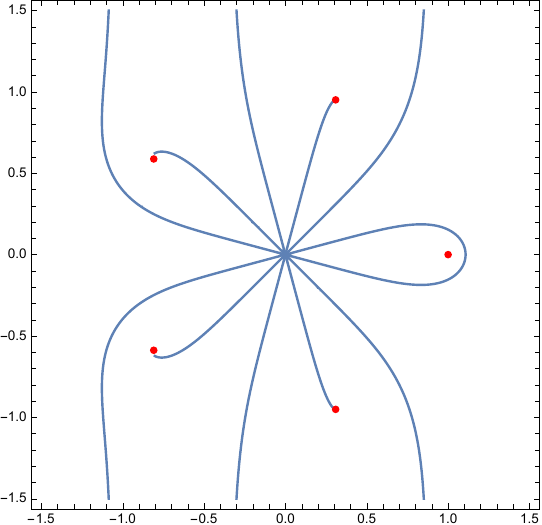}
\caption{$d=8$}
\end{subfigure}
\begin{subfigure}{0.32\textwidth}
\includegraphics[width=5.2cm, height=5.2cm]{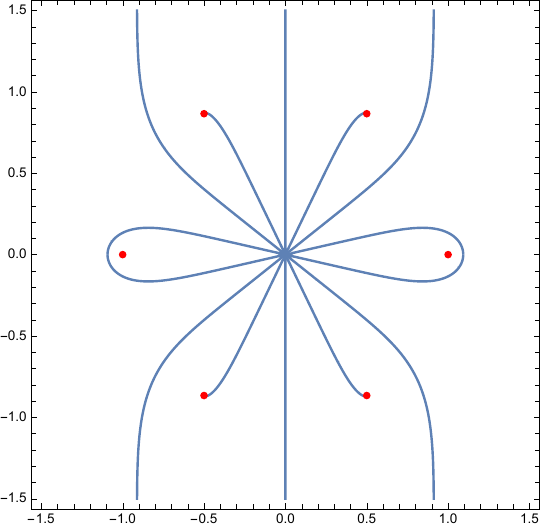}
\caption{$d=9$}
\end{subfigure}
\begin{subfigure}{0.32\textwidth}
\includegraphics[width=5.2cm, height=5.2cm]{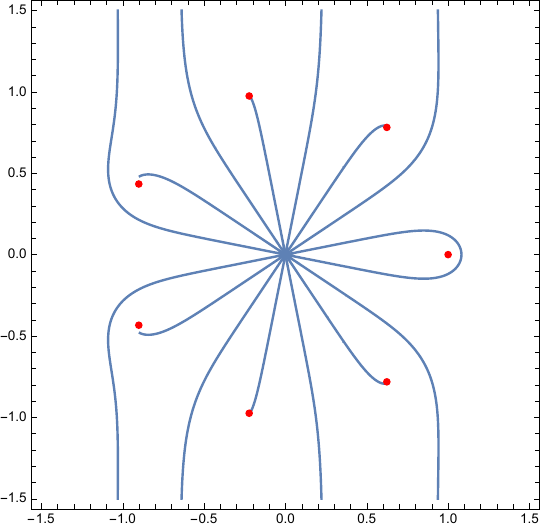}
\caption{$d=10$}
\end{subfigure}
\caption{Numerical plot of the Stokes lines topology for different dimensions. The horizontal axis stands for $\Re(r/R_h)$ and the vertical axis stands for $\Im(r/R_h)$. We denoted the positions of the physical horizon $R_h$ and of the fictitious horizons by red dots.}
\label{fig1}
\end{figure}

Now, we parametrize $r$ as
\begin{equation}
    r = \rho e^{i\theta}
\end{equation}
for $\rho \in \mathbb{R}^+_0$ and $\theta \in [0,2\pi[$. We know that
\begin{equation}
    \Re \left(r^{d-2}\right) = \rho ^{d-2}\cos\left((d-2)\theta\right).
\end{equation}
Equating the expression above to zero yields
\begin{equation}
    \theta =\frac{\pi}{d-2} \left(n + \frac{1}{2}\right)
    \label{75}
\end{equation}
for $1 \le n \le 2d-4$. Thus, we know there are 2$d$-4 Stokes lines emerging from the origin of the complex $r$-plane, all equally distributed and separated by an angle of $\frac{\pi}{d-2}$ radians.
In order to show the topology of Stokes lines, far from the origin of the complex $r$-plane, we display, in figure \ref{fig1}, several numerical plots of the these lines, for different dimensions.

By looking at this figure, we first notice the topology of these lines, near the origin of the complex $r$-plane, is as we predicted. Moreover, for every dimension displayed in the figure, it appears to always exist two Stokes lines emerging from the origin of the complex $r$-plane, encircling the event horizon $R_h$. Furthermore, these lines are always followed, clockwise and counter clockwise, by two unbounded Stokes lines. This feature does not restrict itself to the dimensions displayed in figure \ref{fig1}, being a general feature for every dimension $d \ge 4$. In \cite{Natario}, a good justification of this fact is presented.

Finally, for purposes that will become clear later, we also want to know the topology of the Stokes lines associated with the WKB approximation of the master equation $(\ref{36})$. Although we do not show this here, it should be clear that if we were to set up the WKB formalism around $(\ref{36})$, just as we did in the previous section with equation $(\ref{65})$, we would end up with a WKB approximation analogous to  $(\ref{61})$, with $z$ replaced by $x$. Hence, the Stokes lines associated with this approximation are defined by the condition
\begin{equation}
    \Re\left(x(r)\right) = 0.
    \label{72}
\end{equation}

Integrating equation $(\ref{63})$, yields the asymptotic expansion
\begin{equation}
    z(r) \sim x(r) + \mathcal{C}
    \label{73}
\end{equation}
as $|r| \to +\infty$, for some $\mathcal{C} \in \mathbb{C}$. We set the complex integration constant, defining $x$, to be such that $\mathcal{C} = 0$. Hence, the Stokes lines topologies, associated with the WKB approximations of $(\ref{36})$ and $(\ref{65})$, match, far enough from the origin of the complex $r$-plane. In particular, the two previously mentioned unbounded Stokes lines are also present in the Stokes lines topology of the WKB approximation of $(\ref{36})$, only differing near the origin of the complex $r$-plane. For a more detailed discussion of this point see \cite{Moura:2022gqm}.

\subsection{Boundary conditions}
\noindent

Here, we discuss the boundary conditions, in spatial infinity and in the event horizon $R_h$.

First, we introduce a common issue in the quasinormal mode problem. It should be clear that any method devised to compute quasinormal frequencies will have to make use of the defining boundary conditions $(\ref{20})$ and $(\ref{21})$. Because quasinormal frequencies are complex, gathering information of these boundary conditions amounts to distinguish between an exponentially small and an exponentially large term. Clearly, a numerical approach will face problems with such task. Moreover, any kind of analytical approximate approach will also fail to some extent. Indeed, many lower order terms of the approximation are needed in order to make sense of the exponentially decreasing term, otherwise this term might be much smaller than the approximation error and consequently needs to be disregarded.

In the previous section, this problem was not too worrisome because quasinormal frequencies, in the eikonal limit, are approximately real. However, as we know, this is not true in general.

An elegant solution to half of this issue emerges from the moment we allow $r$ to take complex values and consequently assume an analytic continuation of functions of $r$ to the complex plane. Indeed, we can distinguish the two exponential terms near the event horizon, by computing the respective monodromies around it. These monodromies are non trivial because $x$ has a branch point in the event horizon.

So, we are left with the task of imposing the boundary condition $(\ref{21})$. As it turns out, the solution to this problem also comes with the analytic continuation of functions of $r$ to the complex plane. Indeed, imposing the boundary condition $(\ref{21})$ is equivalent to imposing the boundary condition
\begin{equation}
    \psi(x) \propto e^{-i\omega x}
    \label{71}
\end{equation}
for $|r| \to +\infty$, in any of the two unbounded Stokes lines. Looking at $(\ref{72})$, we know that
\begin{equation}
    |e^{i\omega x}| \sim |e^{-i\omega x}| \sim 1
\end{equation}
in these lines. Thus, imposing the boundary condition $(\ref{71})$, no longer poses a challenge to an approximate analytical method.

Looking at equation $(\ref{73})$, we notice we can rewrite the boundary condition $(\ref{71})$ as
\begin{equation}
    \psi(z) \propto e^{-i\omega z}
    \label{74}
\end{equation}
for $|r| \to +\infty$, in the same Stokes lines. This boundary condition applies to the differential equation $(\ref{64})$ as the latter is simply $(\ref{36})$, rewritten with respect to the independent variable $z$. Finally, the boundary condition $(\ref{74})$ also applies to $(\ref{65})$ and $(\ref{66})$, for both arise from a perturbative approach of $(\ref{64})$. In fact, the differential equation $(\ref{65})$ is the master equation associated with tensor type gravitational perturbations, in the $d$-dimensional Tangherlini black hole space time \cite{Natario}.

\subsection{The monodromy method}
\noindent

The monodromy method uses both solutions, introduced in the previous subsection, to impose the appropriate boundary conditions. The general idea goes as follows:

\begin{itemize}

    \item We pick two closed homotopic contours on the complex $r$-plane. Both these contours enclose only one horizon: the real one $R_h$. Moreover, neither one of them encloses the origin of the complex $r$-plane.
    \item One of these contours, which we name the big contour, seeks to encode information of the boundary condition $(\ref{74})$ on the monodromy of $\psi$, associated with a full loop around it.
    \item The other contour, which we name the small contour, seeks to encode information of the boundary condition $(\ref{20})$ on the monodromy of $\psi$, associated with a full loop around it.
    \item As both contours are homotopic, the monodromy theorem asserts that the respective monodromies must be the same. Thus, equating them hopefully yields a restriction on the values of the quasinormal frequencies $\omega$, from the complex plane to an infinite but countable subset.

\end{itemize}

\subsection{The big contour}
\noindent

In order to build the big contour, we make use of the feature we found, when studying the topology of the Stokes lines, associated with the WKB approximation $(\ref{61})$. More precisely, for every dimension $d$, there will be two Stokes lines, emerging from the origin of the complex $r$-plane, encircling the event horizon $R_h$. Furthermore, these lines are followed, counter clockwise and clockwise, by two unbounded Stokes lines. The big contour will follow these unbounded Stokes lines, reaching the condition $|r| \to +\infty$ twice. There, the contour abandons the Stokes lines and follows an arc shaped path enclosing it.

\begin{figure}[h]
\centering
\includegraphics[width=0.5\textwidth]{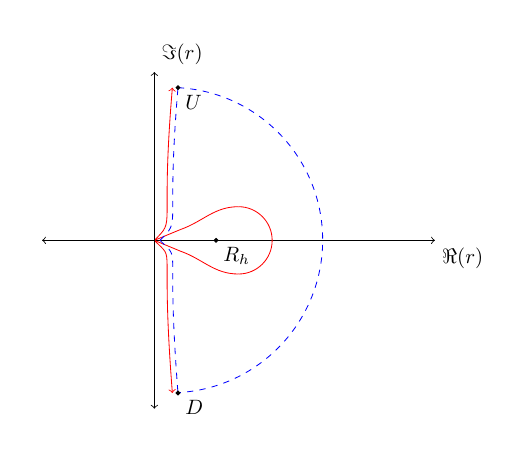}
\caption{Schematic depiction of the big contour, as the blue dashed line. The Stokes lines are depicted as red curves. Naturally, not all Stokes lines are depicted. Furthermore, we marked by $D$ and $U$ the regions where the boundary condition (\ref{74}) may be imposed.}
\label{fig5}
\end{figure}

Overall, the big contour is well represented as depicted in figure \ref{fig5} (we stress that in this figure the proportions may not be right). However, the topology of the big contour is well represented by the blue dashed line for every dimension $d \ge 5$. The boundary condition $(\ref{74})$ is to be imposed in the regions marked by $D$ or $U$.

After defining the big contour, we want to compute the monodromy of $\psi$, associated with a full clockwise loop around it. In order to attain reliable information on how $\psi$ changes around the big contour, we resort to WKB theory. Indeed, we can use $(\ref{61})$ in portions of the big contour, matched with Stokes lines, far enough from the the origin of the complex $r$-plane. Near the origin, the WKB approximation fails as $r=0$ is a singular point of $(\ref{58})$. Thus, we need to analytically solve the differential equations (\ref{65}) and (\ref{66}) in this region.

The full calculation of the monodromy of $\psi$ around the big contour is a long and complex task, full of subtleties and technical details, whose reproduction is outside the scope of this review. We refer the reader to \cite{Moura:2022gqm,Moura:2021nuh}, where it is explained in detail.

\subsection{The small contour}
\noindent

\begin{figure}[H]
\centering
\includegraphics[width=0.5\textwidth]{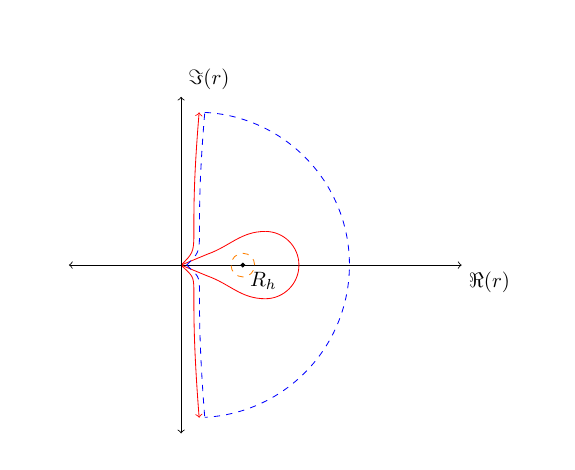}
\caption{Schematic depiction of the small and big contours as the orange and blue dashed lines respectively. The orange contour is to be interpreted as arbitrarily close to $R_H$. The Stokes lines are depicted by red curves. Naturally, not all Stokes lines are depicted.}
\label{fig7}
\end{figure}

The small contour is remarkably simple, when compared to the big contour. Indeed, we build an arbitrarily small closed contour around the event horizon $R_h$. Such contour can be represented as depicted in figure \ref{fig7}. In this contour, we won't need to perturbatively approach the master equation $(\ref{36})$ nor change the independent variable. Indeed, we can simply consider the solution (\ref{ho}) to (\ref{36}), together with the boundary condition $(\ref{20})$, and (as the contour is arbitrarily small) take a Taylor expansion around $r=R_h$. The tortoise (\ref{69}) has a branch point at $r=R_h$, from which we can compute the monodromy of $\psi$ around the small contour. The calculation of the monodromy is therefore considerably simpler when compared to the case of the big contour. Again we refer the reader to \cite{Moura:2022gqm,Moura:2021nuh}, where it can be found in detail.

\subsection{Equating monodromies: the final result}
\noindent

We notice the big contour is homotopic to the small one. Thus, by the monodromy theorem we know the monodromies of $\psi$, associated with the full clockwise loops around the big and the small contours, are equal. Hence, equating the two monodromies we obtain an algebraic condition restraining the possible values of the asymptotic quasinormal frequencies $\omega$, written in terms of the black hole temperature (\ref{154}):
\be
\frac{\omega}{T_\mathcal{H}} = \ln(3) + (2k + 1)\pi i
+ \lambda \left(\frac{4 \pi}{d-3}\right)^2 T_\mathcal{H}^2 \left[\frac{d-3}{d-2} \frac{(2k + 1)}{4}\right]^{\frac{d-1}{d-2}} \Pi_{\textsf{T}}(d) \, \mathrm{e}^{\frac{d-5}{2(d-2)}\pi i}, \label{wtt4}
\ee
with $k \in \mathbb{N}$ and $\Pi_{\textsf{T}}(d)$ given by
\be
\Pi_{\textsf{T}}(d) := \frac{2}{3} \frac{(d-4) \left(d(d-5)+2\right)}{d-1} \sqrt{\pi} \sin\left(\frac{\pi}{d-2}\right) \frac{\Gamma\left(\frac{1}{2(d-2)}\right) \Gamma\left(\frac{d-3}{2(d-2)}\right) }{\left[\Gamma\left(\frac{d-1}{2(d-2)}\right)\right]^2}.
\ee
The $\lambda=0$ limit corresponds to the result obtained in \cite{Natario,motl}: the real part is a universal value $\ln(3)$; the imaginary part depends only on the mode number $k$. In contrast, the $\lambda$ corrections depend strongly on the dimension $d$, both for the real and the imaginary parts. The result we derived here for tensorial perturbations of the Callan-Myers-Perry black hole is also valid for gravitational perturbations of any type (tensorial, vectorial and scalar) of $d$-dimensional Gauss-Bonnet corrected black holes, as shown in \cite{Moura:2022gqm}. In this same reference, and in \cite{Moura:2021nuh}, one can also find the calculations of asymptotic quasinormal modes of test scalar fields.

\paragraph{Acknowledgements}
\noindent
The authors thank the organization committee of the 11th Aegean Summer School that took place in Syros,
Greece, where this article was presented. This work has been supported by Funda\c c\~ao para a Ci\^encia e a Tecnologia under grants IT (UIDB/50008/2020 and UIDP/50008/2020), CAMGSD/IST-ID (UIDB/04459/2020 and UIDP/04459/2020) and projects CERN/FIS-PAR/0023/2019, 2022.08368.PTDC. Jo\~ao Rodrigues is supported by Funda\c c\~ao para a Ci\^encia e a Tecnologia through the doctoral fellowship UI/BD/151499/2021.

\end{document}